\documentclass{sigchi}
\pdfoutput=1

\CopyrightYear{2020}

\conferenceinfo{CHI'20,}{April  25--30, 2020, Honolulu, HI, USA}
\acmPrice{\$15.00}

\toappear{}

\usepackage{balance}       
\usepackage{graphics}      
\usepackage[T1]{fontenc}   
\usepackage{txfonts}
\usepackage{mathptmx}
\usepackage{color}
\usepackage{booktabs}
\usepackage{textcomp}
\usepackage{subcaption}

\usepackage{graphicx}
\usepackage{multirow}
\usepackage{booktabs}

\usepackage{microtype}        
\usepackage{ccicons}          

\usepackage{todonotes}

\usepackage{hyperref}
\hypersetup{pdflang={en-US},pdftex}

\def\plaintitle{The Newspaper Navigator Dataset:  \\ Extracting And Analyzing Visual Content from 16 Million Historic Newspaper Pages in Chronicling America}
\def\emptyauthor{}
\def\plainkeywords{Information Extraction; Image Segmentation; Historic Newspapers; \href{https://chroniclingamerica.loc.gov/}{Chronicling America}; digital humanities}

\makeatletter
\def\url@leostyle{%
  \@ifundefined{selectfont}{
    \def\UrlFont{\sf}
  }{
    \def\UrlFont{\small\bf\ttfamily}
  }}
\makeatother
\def\pprw{8.5in}
\def\pprh{11in}

\definecolor{linkColor}{RGB}{6,125,233}
\hypersetup{%
  pdftitle={\plaintitle},
  pdfauthor={\emptyauthor},
  pdfkeywords={\plainkeywords},
  pdfdisplaydoctitle=true, 
  bookmarksnumbered,
  pdfstartview={FitH},
  colorlinks,
  citecolor=black,
  filecolor=black,
  linkcolor=black,
  urlcolor=linkColor,
  breaklinks=true,
  hypertexnames=false
}

\begin{document}

\title{\plaintitle}

\numberofauthors{4}
\author{%
  \alignauthor{Benjamin Lee\thanks{Work conducted while an Innovator-in-Residence at the Library of Congress and Ph.D. student in the Paul G. Allen School for Computer Science and Engineering at the University of Washington.}
  \\
    \affaddr{University of Washington}\\ \affaddr{Library of Congress}\\
\href{mailto:bcgl@cs.washington.edu}{bcgl@cs.washington.edu}}\\
  \alignauthor{Jaime Mears\\
    \affaddr{LC Labs}\\
    \affaddr{Library of Congress}}\\
  \alignauthor{Eileen Jakeway\\
    \affaddr{LC Labs}\\
    \affaddr{Library of Congress}}\\
  \alignauthor{Meghan Ferriter\\
    \affaddr{LC Labs}\\ 
    \affaddr{Library of Congress}}\\
  \alignauthor{Chris Adams\\
    \affaddr{IT Design \& Development}
    \affaddr{Library of Congress}}\\
  \alignauthor{Nathan Yarasavage\\
    \affaddr{National Digital Newspaper Program}\\
    \affaddr{Library of Congress}}\\
  \alignauthor{Deborah Thomas\\
    \affaddr{National Digital Newspaper Program}\\
    \affaddr{Library of Congress}}\\
  \alignauthor{Kate Zwaard\\
    \affaddr{Digital Strategy \& LC Labs}\\
    \affaddr{Library of Congress}}\\
  \alignauthor{Daniel Weld\\
    \affaddr{University of Washington}}\\
}

\maketitle

\section{Abstract}
Chronicling America is a product of the National Digital Newspaper Program, a partnership between the Library of Congress and the National Endowment for the Humanities to digitize historic newspapers. Over 16 million pages of historic American newspapers have been digitized for Chronicling America to date, complete with high-resolution images and machine-readable METS/ALTO OCR. Of considerable interest to Chronicling America users is a semantified corpus, complete with extracted visual content and headlines. To accomplish this, we introduce a visual content recognition model trained on bounding box annotations of photographs, illustrations, maps, comics, and editorial cartoons collected as part of the Library of Congress's Beyond Words crowdsourcing initiative and augmented with additional annotations including those of headlines and advertisements. We describe our pipeline that utilizes this deep learning model to extract 7 classes of visual content: headlines, photographs, illustrations, maps, comics, editorial cartoons, and advertisements, complete with textual content such as captions derived from the METS/ALTO OCR, as well as image embeddings for fast image similarity querying. We report the results of running the pipeline on 16.3 million pages from the Chronicling America corpus and describe the resulting Newspaper Navigator dataset, the largest dataset of extracted visual content from historic newspapers ever produced. The Newspaper Navigator dataset, finetuned visual content recognition model, and all source code are placed in the public domain for unrestricted re-use.

 \section{Introduction}
Chronicling America, an initiative of the National Digital Newspaper Program - itself a partnership of the Library of Congress and the National Endowment for the Humanities - is an invaluable resource for academic, local, and public historians; educators and students; genealogists; journalists; and members of the public to explore American history through the uniquely rich content preserved within historic local newspapers. Over 16 million pages of newspapers published between 1789 to 1963 are publicly available online through a search portal, as well as via a public API. Among the page-level data are 400 DPI images, as well as METS/ALTO OCR, a standard maintained by the Library of Congress that includes text localization \cite{about_chronam}.  

\begin{figure*}[t!]
    \centering
    \begin{subfigure}[t]{0.5\textwidth}
        \centering
        \includegraphics[height=2.3in]{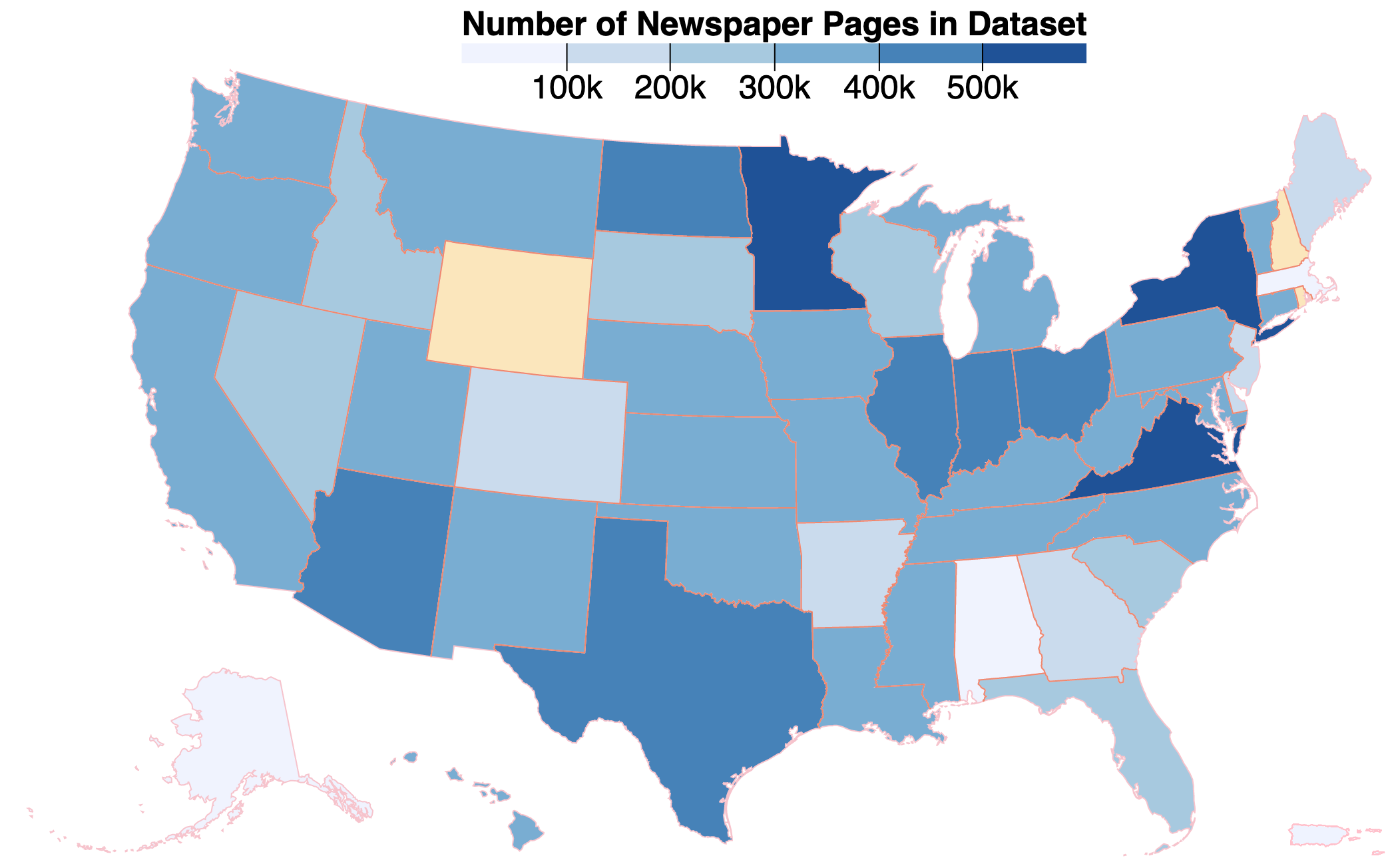}
        \caption{State-level choropleth map.}
    \end{subfigure}%
    ~ 
    \begin{subfigure}[t]{0.5\textwidth}
        \centering
        \includegraphics[height=2.3in]{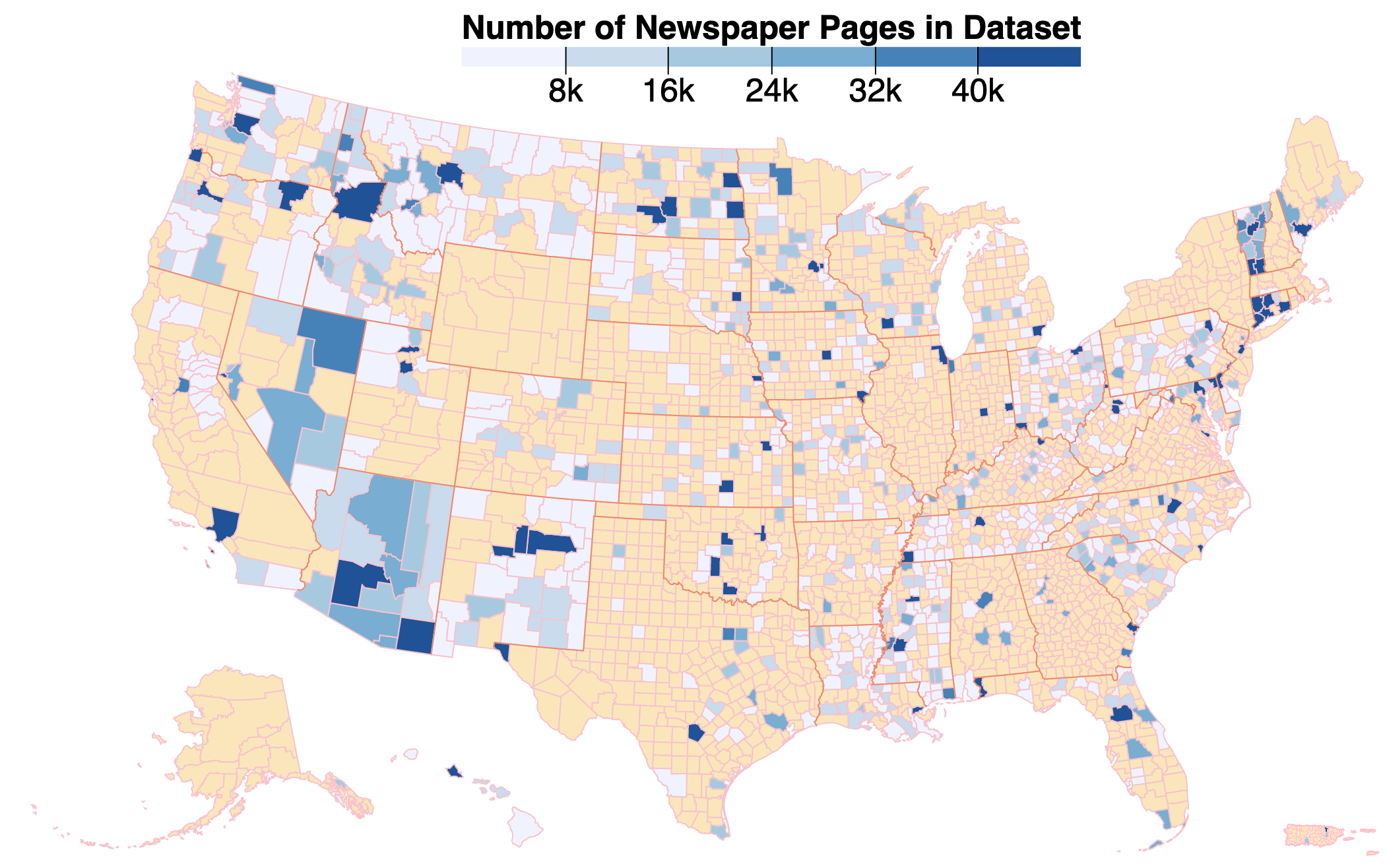}
        \caption{County-level choropleth map.}
    \end{subfigure}
    \caption{Choropleth maps at the state and county level showing the geographic coverage of the 16.3 million Chronicling America historic newspaper pages included in the Newspaper Navigator dataset. Yellow coloring indicates no pages cover the corresponding region. Puerto Rico is pictured in the bottom-right of each map.}~\label{fig:choropleth}
\end{figure*}

\begin{figure}[t!]
\centering
  \includegraphics[width=1.0\columnwidth]{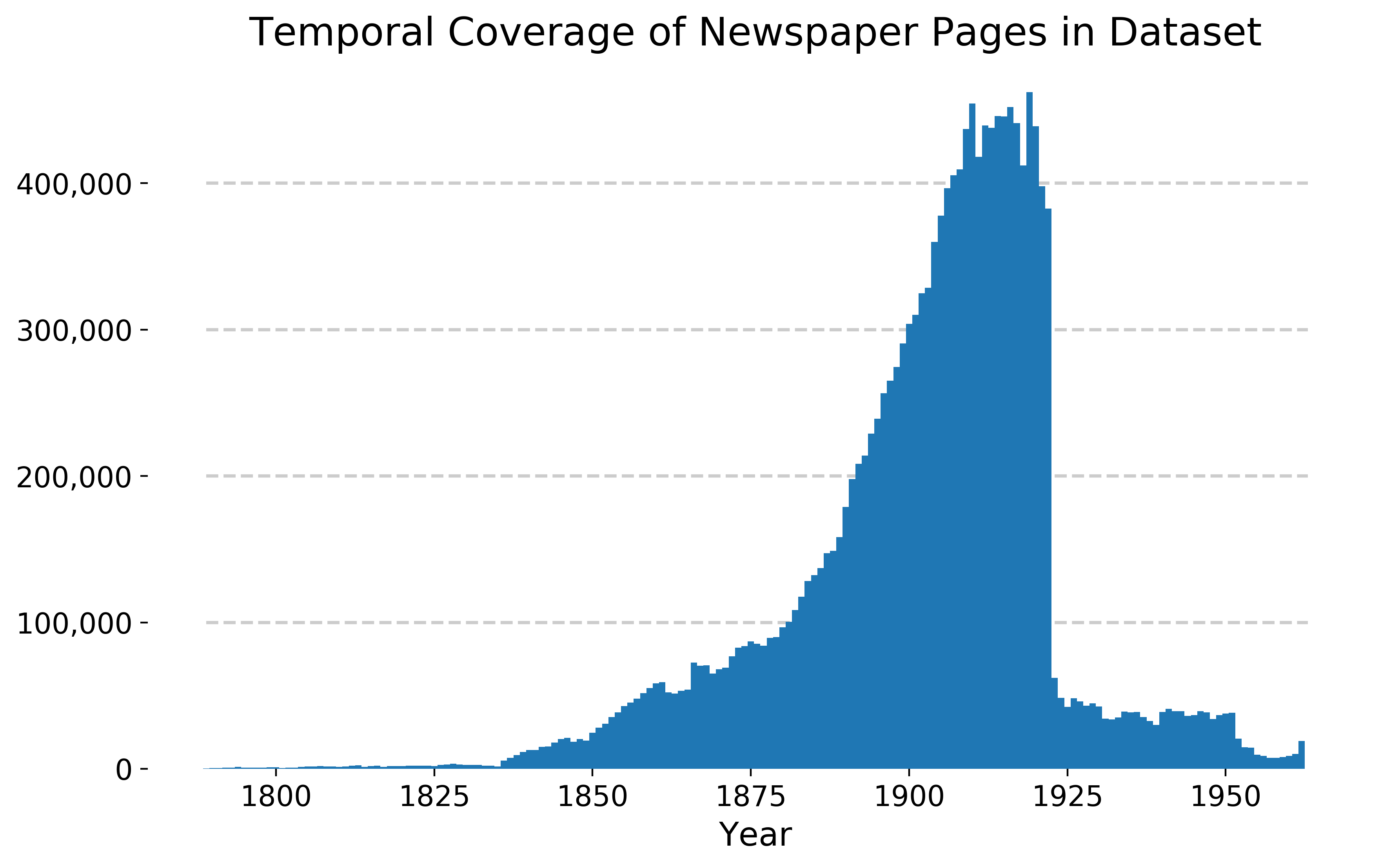}
  \caption{A histogram showing the temporal coverage of the 16.3 million Chronicling America historic newspaper pages included in the Newspaper Navigator dataset. A temporal cutoff of 1922 was used for Chronicling America newspaper digitization until 2016, explaining the corresponding dropoff in temporal coverage.}~\label{fig:dates}
\end{figure}

The 16.3 million Chronicling America pages included in the Newspaper Navigator cover 174 years of American history, inclusive of 47 states, Washington, D.C., and Puerto Rico. In Figure \ref{fig:choropleth}, we show choropleth maps displaying the geographic coverage of the 16.3 million Chronicling America newspaper pages included in the Newspaper Navigator dataset. In Figure \ref{fig:dates}, we show the temporal coverage of these pages. The coverage reflects the selection process for determining which newspapers to include in Chronicling America; for an in-depth examination, please refer to \cite{quoth,chronicling_america_guidelines}. The selection process should be considered in the methodology of any research performed using the Newspaper Navigator dataset.

While the images and OCR in Chronicling America provide a wealth of information, users interested in extracted visual content, including headlines, are currently restricted to general keyword searches or manual searches over individual pages in Chronicling America. For example, staff at the Library of Congress have produced a collection of Civil War maps in historic newspapers to date, but the collection is far from complete due to the difficulty of manually searching over the hundreds of thousands of Chronicling America pages from 1861 to 1865 \cite{civil_war}. A complete dataset would be of immense value to historians of the Civil War. Likewise, collecting all of the comic strips from newspapers published in the early 20th century would provide comic researchers with a corpus of unprecedented scale. In addition, users currently have no reliable method of determining what disambiguated articles appear on each page, presenting challenges for natural language processing (NLP) approaches to studying the corpus. A dataset of extracted headlines not only gives researchers insight into the individual articles that appear on each page but also enables users to ask questions such as, ``Which news topics appeared above the fold versus below the fold in what newspapers?'' Indeed, the digital humanities questions that could be asked with such a dataset abound.  And yet, the possibilities extend beyond the digital humanities to include public history, creative computing, educational use within the classroom, and public engagement with the Library of Congress's collections.

 To begin the construction of larger datasets of visual content within Chronicling America and to engage the American public, the Library of Congress Labs launched a crowdsourcing initiative called Beyond Words\footnote{\url{https://labs.loc.gov/work/experiments/beyond-words/}} in 2017.  With this initiative, volunteers were asked to draw bounding boxes around photographs, illustrations, comics, editorial cartoons, and maps in World War 1-era newspapers in Chronicling America; they were also asked to transcribe captions by correcting the OCR within each bounding box annotation, as well as record the content creator. Approximately 10,000 verified Beyond Words annotations have been collected to date.
 
Our research builds on the crowdsourced Beyond Words annotations by utilizing the bounding boxes drawn around photographs, illustrations, comics, editorial cartoons, and maps, as well as additional annotations including ones marking headlines and advertisements, to finetune a pre-trained Faster-RCNN implementation from Detectron2's Model Zoo \cite{ren_faster_2015, detectron2}.  Our visual content recognition model predicts bounding boxes around these 7 different classes of visual content in historic newspapers. This paper presents our work training this visual content recognition model and constructing a pipeline for automating the identification of this visual content in Chronicling America newspaper pages.  Drawing inspiration from the Beyond Words workflow, we extract corresponding textual content such as headlines and captions by identifying text from the METS/ALTO OCR that falls within each predicted bounding box.   This method is effective at captioning because Beyond Words volunteers were asked to include captions and relevant textual content within their bounding box annotations. Lastly, in order to enable fast similarity querying for search and recommendation tasks, we generate image embeddings for the extracted visual content using ResNet-18 and ResNet-50 models pre-trained on ImageNet. This resulting dataset, which we call the Newspaper Navigator dataset, is the largest collection of extracted visual content from historic newspapers ever produced.

Our contributions are as follows:
\begin{enumerate}
  \setlength{\itemsep}{0pt}
    \item We present a publicly available pipeline for extracting visual and textual content from historic newspaper pages, designed to run at scale over terabytes of image data. Visual content categories include headlines, photographs, illustrations, maps, comics, editorial cartoons, and advertisements. 
    \item We release into the public domain a finetuned Faster-RCNN model for this task that achieves 63.4\% bounding box mean average precision (mAP)\footnote{Mean average precision is the standard metric used for benchmarking object detection models, incorporating intersection over union to assess precision and recall. We describe the metric in more detail in Section \ref{sec:training}.} on a validation set of World War 1-era Chronicling America pages.
    \item We present the Newswpaper Navigator dataset, a new public dataset of extracted headlines and visual content, as well as corresponding textual content such as titles and captions, produced by running the pipeline over 16.3 million historic newspaper pages in Chronicling America.  This corpus represents the largest dataset of its kind ever produced.
\end{enumerate}

\section{Related Work}\label{sec:related}

\subsection{Corpora \& Datasets}
Over the past 15 years, efforts across the world to digitize historic newspapers have been remarkably successful \cite{digitalturn}. In addition to Chronicling America, examples of large repositories of digitized newspapers include Trove \cite{trove}, Europeana \cite{pekarek_europeana_2012, willems_europeana_2015}, Delpher \cite{delpher}, The British Newspaper Archive \cite{british}, OurDigitalWorld \cite{ourdigitalworld}, Papers Past \cite{paperspast}, NewspaperSG \cite{newspapersg}, newspapers.com \cite{newspapersdotcom} and Google Newspaper Search \cite{chaudhury_google_2009}. The availability of newspapers at the scale of millions of digitized pages has inspired the construction of datasets for supervised learning tasks related to digitized newspapers. In addition to Beyond Words, datasets for historic newspaper recognition include the National Library of Luxembourg's historic newspaper datasets \cite{BnL} that include segmented articles and advertisements; CHRONIC, a dataset of 452,543 images in historic Dutch newspapers \cite{chronic}; and Europeana's SIAMESET, a dataset of 426,777 advertisements in historic Dutch newspapers \cite{siameset}. Datasets for machine learning tasks with historical documents include READ-BAD \cite{readbad} and DIVA-HisDB \cite{diva-hisdb}. However, all of these datasets are designed to serve as training sets rather than as comprehensive datasets of extracted content from full corpora. Our work instead seeks to use the Beyond Words dataset to train a visual content recognition model in order to process the visual content in the Chronicling America corpus comprising 16+ million historic newspaper pages.

\subsection{Visual Content Extraction}

Other researchers have built tools and pipelines for extracting and analyzing visual content from historic documents, including newspapers, using deep learning.\footnote{For approaches to historic document classification that do not utilize deep learning, see for example \cite{lee_ITS}.} PageNet utilizes a Fully Convolutional Network for pixel-wise page boundary extraction for historic documents \cite{pagenet}. dhSegment is a deep learning framework for historical document processing, including pixel-wise segmentation and extraction tasks \cite{dhsegment}. Liebl and Burghardt benchmarked 11 different deep learning backbones for the pixel-wise segmentation of historic newspapers, including the separation of layout features such as text and tables \cite{liebl2020evaluation}. The AIDA collaboration at the University of Nebraska-Lincoln has applied deep learning techniques to newspaper corpora including Chronicling America and the Burney Collection of British Newspapers \cite{lorang_patterns,lorang_application,lorang_using} for tasks such as poetic content recognition \cite{DLIB_AIDA, AAAI_AIDA}, as well as visual content recognition using dhSegment \cite{UNL_report}. Instead of a pixel-wise approach, we instead utilize bounding boxes, resulting in higher performance. In addition, our pipeline recognizes 7 different classes of visual content (headlines, photographs, illustrations, maps, comics, editorial cartoons, and advertisements), extracts corresponding OCR, and generates image embeddings. Lastly, we deploy our visual content recognition pipeline at scale.

\subsection{Article Disambiguation}
Article disambiguation for historic newspaper pages has long been of interest to researchers.  Groups that have studied this task include the IMPRESSO project \cite{impresso}, NewsEye project \cite{newseye}, and Google Newspaper Search \cite{chaudhury_google_2009}.\footnote{Related work has focused on content segmentation for books \cite{bamman_books}.}  Of particular note is the approach taken by Google Newspaper Search, which extracted headline blocks using OCR font size and area-perimeter ratio as features and utilized the extracted headlines to segment each page into individual articles \cite{chaudhury_google_2009}.\footnote{To our knowledge, the extraction and classification of visual content was
outside of the scope of the project.} We, too, focus on headline extraction because it serves as its own form of article disambiguation. However, unlike previous approaches, we treat headline extraction as a \textit{visual} task at the image level, rather than a \textit{textual} task at the OCR level. Our novel approach is to leverage the visual distinctiveness of headlines on the newspaper pages and train a classifier to predict bounding boxes around headlines on the page. The headline text within each bounding box is then extracted from the underlying METS/ALTO OCR.

Lastly, it should be noted that proper article disambiguation requires the ability to filter out text from advertisements due to the ubiquity of advertisements. As with headlines, we treat advertisement identification as a visual task rather than a textual task because the advertisements are so naturally identified by their visual features. Because our visual content recognition model robustly identifies advertisements, we are able to disambiguate newspaper text from advertisement text. 

\subsection{Image Embeddings for Cultural Heritage Collections}
In recent years, researchers have utilized image embeddings for visualizing and exploring visual content in cultural heritage collections. The Yale Digital Humanities Lab's PixPlot interface \cite{pixplot} and the National Neighbors project \cite{nationalneighbors} utilize Inception v3 embeddings \cite{inceptionv3}. Google Arts \& Culture's t-SNE Map utilizes embeddings produced by the  Google search pipeline \cite{tsnemap}. The Norwegian National Museum's Principal Components project \cite{principalcomponents} uses finetuned Caffe image embeddings \cite{caffe}. Olivia Vane utilizes VGG-16 embeddings to visualizing the Royal Photographic Society Collection \cite{vane}. Likewise, Brian Foo has created a visualization of The American Museum of Natural History's image collection \cite{amnh} using VGG-16 embeddings \cite{vgg}. Refik Anadol uses embeddings to visualize the SALT Research collection \cite{anadol}. Regarding visual content in historic newspapers in particular, Wevers and Smits have utilized Inception v3 embeddings to perform large-scale analysis of the CHRONIC and SIAMESET datasets derived from historic Dutch newspapers \cite{visualturn}. Their work includes the deployment of SIAMESE, a recommender system for advertisements in historic newspapers, as well as an analysis of training a new classification layer on top of the Inception embeddings to predict according to custom categories \cite{visualturn}.

Indeed, in addition to supporting visualizations of latent spaces that capture semantic similarity, image embeddings are desirable for visual search and recommendation tasks due to the ability to perform fast similarity querying with them. Using ResNet-18 and ResNet-50 \cite{resnet} models pre-trained on ImageNet, we generate image embeddings for the extracted visual content, which are included in the Newspaper Navigator dataset in order to support a range of visual search and recommendation tasks for the Chronicling America corpus.

\section{Code}

All code discussed in this paper can be found in the public GitHub repository \url{https://github.com/LibraryOfCongress/newspaper-navigator} and is open source, placed in the public domain for unrestricted re-use.  In addition, included in the repository are the finetuned visual content recognition model, the training set on which the model was finetuned, a Jupyter notebook for experimenting with the visual content recognition model, and a slideshow of predictions.

\section{Constructing the Training Set}

\subsection{Repurposing the Beyond Words Annotations}

To create a training set for our visual content recognition model, we repurposed the publicly available annotations of photographs, illustrations, maps, comics, and editorial cartoons derived from Beyond Words, a crowdsourcing initiative launched by the Library of Congress to engage the American public with the visual content in World War 1-era newspapers in Chronicling America. The Beyond Words platform itself was built using Scribe \cite{scribe}. The crowdsourcing workflow consisted of three different tasks that volunteers could choose to perform: 
\begin{enumerate}
\itemsep0em 
    \item \textit{Mark}, in which users were asked to ``draw a rectangle around each unmarked illustration or photograph excluding those in advertisements [and] enclose any caption or text describing the picture and the illustrator or photographer'' \cite{mark}.
    \item \textit{Transcribe}, in which users were asked to correct the OCR of the caption for each marked illustration or photograph, transcribe the author's name, and note the category (``Editorial Cartoon,'' ``Comics/Cartoon,'' ``Illustration,'' ``Photograph,'' or ``Map'') \cite{transcribe}.
    \item \textit{Verify}, in which users were asked to select the transcription of another volunteer that most closely matches the printed caption.  Users were also able to filter out bad regions or provide their own transcriptions in the event that neither transcription was of good quality \cite{verify}.
\end{enumerate}
Up to 6 different individuals may have interacted with each annotation during this process.  The annotation required achieving at least 51\% agreement with volunteers at the \textit{Transcribe} and \textit{Verify} steps.

In order to finetune the visual content recognition model, it was first necessary to reformat the crowdsourced Beyond Words annotations into a proper data format for training a deep learning model.  We chose the Common Objects in Context (COCO) dataset format \cite{coco}, a standard data format for object detection, segmentation, and captioning tasks adopted by Facebook AI Research's Detectron2 deep learning platform for object detection \cite{detectron2}. 

The verified Beyond Words annotations used as training data were downloaded from the Beyond Words public website on December 1, 2019. To convert the JSON file available for download into a deep learning training set, we wrote a Python script to pull down the Chronicling America newspaper images utilized by Beyond Words and format the annotations according to the COCO standard. The script is available in the Newspaper Navigator GitHub repository.

We reiterate that the instructions for the ``Mark'' step asked users to ``enclose any caption or text describing the picture and the illustrator or photographer'' \cite{mark}; therefore, a model trained on these annotations learns to include relevant text within the bounding boxes for visual content, which can then be extracted from the corresponding METS/ALTO OCR in an automated fashion.

\subsection{Adding Annotations}

Because headlines and advertisements were not included in the Beyond Words workflow, we added annotations for headlines and advertisements for all images in the dataset. These annotations are not verified, as each page was annotated by only one person. In addition, due to the low number of annotated maps in the Beyond Words data (79 in total), we added annotations of 122 pages containing maps, which were retrieved by performing a keyword search of ``map'' on the Chronicling America search portal restricted to the years 1914-1918. We then downloaded the pages on which we identified maps, and we annotated all 7 categories of visual content on each page. Like the headline and advertisement annotations, these annotations are not verified.

\subsection{Training Set Statistics}
The augmented Beyond Words dataset in COCO format can be found in the Newspaper Navigator repository and is available for unrestricted re-use in the public domain.  The dataset contains $3,559$ World War 1-era Chronicling America pages with $48,409$ annotations in total.  The category breakdown of annotations appears in Table \ref{tab:breakdown}.

\begin{table}
\centering
\begin{tabular}{cc} 
\multicolumn{2}{c}{\textbf{Training/Validation Set Statistics}} \\
\toprule
\centering
\textbf{Category} & \textbf{Count} \\
\midrule
Photograph & 4,254 \\
\hline
Illustration & 1,048 \\
\hline
Map & 215 \\
\hline
Comic/Cartoon & 1,150 \\
\hline
Editorial Cartoon & 293 \\
\hline
Headline & 27,868 \\
\hline
Advertisement & 13,581 \\
\hline
\textit{Total} & 48,409 \\
\bottomrule
\end{tabular}
\caption{A table showing a breakdown of content for the 7 different classes in the training/validation dataset produced using the Beyond Words bounding box annotations, augmented with additional annotations.}\label{tab:breakdown}
\end{table}

\section{Training the Visual Content Recognition Model}\label{sec:training}
To train a visual content recognition model for identifying the 7 classes of different newspaper content, we chose to finetune a pre-trained Faster-RCNN object detection model from Detectron2's Model Zoo using Detectron2 \cite{detectron2}  and PyTorch \cite{pytorch}. Because model inference was the bottleneck on runtime in our pipeline, we chose the Faster-RCNN R50-FPN backbone, the fastest such backbone according to inference time. Though we could have utilized the highest performing Faster-RCNN backbone, which achieved approximately 5\% higher mean average precision on the COCO \cite{coco} pre-training task at the expense of 2.5x the inference time, qualitative evaluation of predictions with the finetuned R50-FPN backbone indicated that the model was performing sufficiently for our purposes. Furthermore, we conjecture that the performance of our visual content recognition model is limited by noise in the training data, rather than model architecture and selection, for two reasons. First, the ground-truth Beyond Words labels were not complete because volunteers were only required to draw one bounding box per page (though more could be added). Second, there was non-trivial disagreement between Beyond Words annotators for the bounding box marking task due to the heterogeneity of visual content layouts and the resulting ambiguities in the annotation task.\footnote{In regard to the accuracy of the annotations, it is worth noting that Beyond Words was launched as an experiment; consequently, there were no interventions in workflow or community management after its launch, and the accuracy of the resulting annotations should be assessed accordingly.}

\begin{table}
\centering
\begin{tabular}{ ccc }
\multicolumn{3}{c}{\textbf{Performance (Validation)}} \\
\toprule
\centering
\textbf{Category} & \centering \textbf{AP} & \textbf{\# in Val. Set} \\
\midrule
Photograph & 61.6\% & 879 \\
\hline
Illustration & 30.9\% & 206 \\
\hline
Map & 69.5\% & 34 \\
\hline
Comic/Cartoon & 65.6\% & 211 \\
\hline
Editorial Cartoon & 63.0\% & 54 \\
\hline
Headline & 74.3\% & 5,689 \\
\hline
Advertisement & 78.7\% & 2,858 \\
\hline
Averaged (mAP) & 63.4\% & N/A \\
\hline
One Class & 75.1\% & 9,931 \\
\bottomrule
\end{tabular}
\caption{A table showing the average precision (AP) on validation data for the finetuned visual content recognition model on the different categories of content, as well as the number of instances of each category in the validation set. The \textit{Averaged} row includes the mean average precision across the 7 classes. The \textit{One Class} row is computed by combining all visual content into one class and computing average precision using the single class.  This captures how much error is introduced by the detection of visual content versus the classification.}\label{tab:model_performance}
\end{table}

\begin{figure*}
\centering
  \includegraphics[width=2.0\columnwidth]{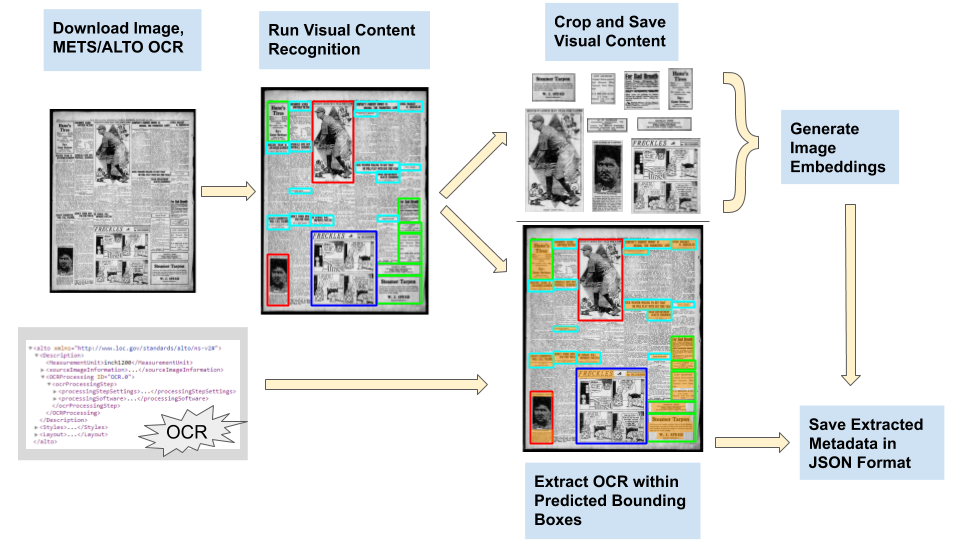}
  \caption{A diagram showing the steps of our pipeline. }~\label{fig:pipeline}
\end{figure*}

All finetuning was performed using PyTorch \cite{pytorch} on a g4dn.2xlarge Amazon EC2 instance with a single NVIDIA T4 GPU. Finetuning the R50-FPN backbone was evaluated on a held-out validation set according to an 80\%-20\% split; the JSON files containing the training and validation splits are available for download with the GitHub repository. We used the following hyperparameters: a base learning rate of  $0.00025$, a batch size of $8$, and $64$ proposals per image. \texttt{RESIZE\_SHORTEST\_EDGE} and \texttt{RANDOM\_FLIP} were utilized as data augmentation techniques.\footnote{These are the only two data augmentation methods currently supported by Detectron2.} Using early stopping, the model was finetuned for 77 epochs, which required 17 hours of runtime on the NVIDIA T4 GPU.  The model weights file is publicly available and can be found in the GitHub repository for this project.

We report a mean average precision on the validation set of 63.4\%; average precision for each category, as well as the number of validation instances in each category, can be found in Table \ref{tab:model_performance}. We chose average precision as our evaluation metric because it is the standard metric utilized in the computer vision community for benchmarking object detection tasks.  Given a fixed intersection over union (IoU) threshold to evaluate whether a prediction is correct, average precision is computed by sorting all classifications according to prediction score, generating the corresponding precision-recall curve, and modifying it by drawing the smallest-area curve containing it that is also monotonically decreasing. According to the COCO standard, average precision is then computed by averaging the precision interpolated over 101 different recall values and 10 IoU thresholds from 50\% to 95\%. For our calculations, we utilized all predictions with confidence scores greater than 0.05 and discarded predictions with confidence scores below this threshold.\footnote{A confidence score of 0.05 is the default threshold cut for retaining predictions in Detectron2.}

\section{The Pipeline}

\subsection{Building the Manifest}

In order to create a full index of digitized pages for the pipeline to process, we used a forked version of the AIDA collaboration's \texttt{chronam-get-images} repository\footnote{\url{https://github.com/bcglee/chronam-get-images}} to generate a manifest for each newspaper batch consisting of filepaths for each page in the batch.\footnote{More information on the batches can be found at \url{https://chroniclingamerica.loc.gov/batches}.} Manifests consisting of 16,368,424 Chronicling America pages were compiled in total on March 17, 2020. 

\subsection{Steps of the Pipeline}
In Figure \ref{fig:pipeline}, we present a diagram showing the pipeline workflow. Each manifest was processed in series by our pipeline. The pipeline code consists of six distinct steps:
\begin{enumerate}
\itemsep0em 
    \item \textit{Downloading the image and METS/ALTO XML for each page and downsampling the image by a factor of 6 to produce a lower resolution JPEG.} Downsampling was performed to reduce I/O and memory consumption, as well as to avoid the overhead introduced by the downsampling that Detectron2 would have to perform before each forward pass during model inference. This step was run in parallel across all 48 CPU cores on each EC2 instance. The files were pulled down directly from the Library of Congress's public AWS S3 buckets.
    \item \textit{Running the visual content recognition model inference on each image to produce bounding box predictions complete with coordinates, predicted classes, and confidence scores.} This step was run in parallel across all 4 GPUs on each EC2 instance. Predictions with confidence scores greater than 0.05 were saved.  We chose to save predictions with low confidence scores in order to allow a user to select a threshold cut based on the user's ideal tradeoff between precision and recall.
    \item \textit{Extracting the OCR within each predicting bounding box.} This step required parsing the METS/ALTO XML and was run in parallel across all 48 CPU cores on each EC2 instance.
    \item \textit{Cropping and saving the extracted visual content as downsampled JPEGs (for all classes other than headlines).} This step was run in parallel across all 48 CPU cores on each EC2 instance.
    \item \textit{Generating ResNet-18 and ResNet-50 embeddings for the cropped and saved images with confidence scores of greater than or equal to 0.5.} This step was implemented using a forked version of img2vec\footnote{\url{https://github.com/bcglee/img2vec}} \cite{img2vec_2019}. This step was run in parallel across all 4 GPUs on each EC2 instance. The ResNet-18 and ResNet-50 embeddings were extracted from the penultimate layer of each respective architecture after being trained on ImageNet (the models themselves were downloaded from \texttt{torchvision.models} in PyTorch \cite{pytorch}). The 2,048-dimensional ResNet-50 embeddings were selected due to ResNet-50's high performance and fast inference time relative to other image recognition models \cite{benchmark}. The 512-dimensional ResNet-18 embeddings were also generated due to their lower dimensionality, enabling faster computation for search and recommendation tasks.
    \item \textit{Saving the extracted metadata and cropped images.} The format of the saved metadata is described in the next section.
\end{enumerate}

\subsection{Running the Pipeline at Scale}

All pipeline processing was performed on 2 g4dn.12xlarge Amazon AWS EC2 instances, each with 48 Intel Cascade Lake vCPUs and 4 NVIDIA T4 GPUs. All pipeline code was written in Python 3.  In total, the pipeline successfully processed 16,368,041 pages ($99.998\%$) in 19 days of wall-clock time.  The manifests of pages that were successfully processed, as well as the 383 pages that failed, can be found in the Newspaper Navigator GitHub Repository.

\begin{table}
\centering
\begin{tabular}{c@{\qquad}ccc}
  \multicolumn{4}{c}{\normalsize{\textbf{Newspaper Navigator Dataset Statistics}}} \\
  \toprule
  \multirow{2}{*}{\raisebox{-\heavyrulewidth}{\normalsize{\textbf{Category}}}} & \multicolumn{3}{c}{\normalsize{\textbf{Count $\mathbb{\ge}$ Confidence Score}}} \\
  \cmidrule{2-4}
  & $\ge$~0.9 & $\ge$~0.7 & $\ge$~0.5 \\
  \midrule
Photograph & 1.59 $\times 10^6$ & 2.63 $\times 10^6$ & 3.29 $\times 10^6$  \\
\hline
Illustration & 8.15 $\times 10^5$ & 2.52 $\times 10^6$ & 4.36 $\times 10^6$  \\
\hline
Map & 2.07 $\times 10^5$ & 4.59 $\times 10^5$ & 7.54 $\times 10^5$  \\
\hline
Comic/Cartoon & 5.35 $\times 10^5$ & 1.23 $\times 10^6$ & 2.06 $\times 10^6$ \\
\hline
Editorial Cartoon & 2.09 $\times 10^5$ & 6.67 $\times 10^5$ & 1.27 $\times 10^6$  \\
\hline
Headline & 3.44 $\times 10^7$ & 5.37 $\times 10^7$ & 6.95 $\times 10^7$  \\
\hline
Advertisement & 6.42 $\times 10^7$ & 9.48 $\times 10^7$ & 1.17 $\times 10^8$ \\
\hline
\textit{Total} & 1.02 $\times 10^8$ & 1.56 $\times 10^8$ & 1.98 $\times 10^8$ \\
  \bottomrule
\end{tabular}
\caption{A table showing a breakdown of extracted content in the Newspaper Navigator dataset. Three different cuts on confidence score are presented to show the effect of the cut choice on the resulting dataset when favoring precision or recall.}\label{tab:breakdownextracted}
\end{table}

\section{The Newspaper Navigator Dataset}

\subsection{Statistics \& Visualizations}

A statistical breakdown of extracted content in the Newspaper Navigator dataset is presented in Table \ref{tab:breakdownextracted}.  Because the choice of threshold cut on confidence score affects the number of resulting visual content in the Newspaper Navigator dataset, we include statistics for three different threshold cuts of 0.5, 0.7, and 0.9.

\begin{figure*}[t!]
    \centering
    \begin{subfigure}[t]{0.5\textwidth}
        \centering
        \includegraphics[height=5in]{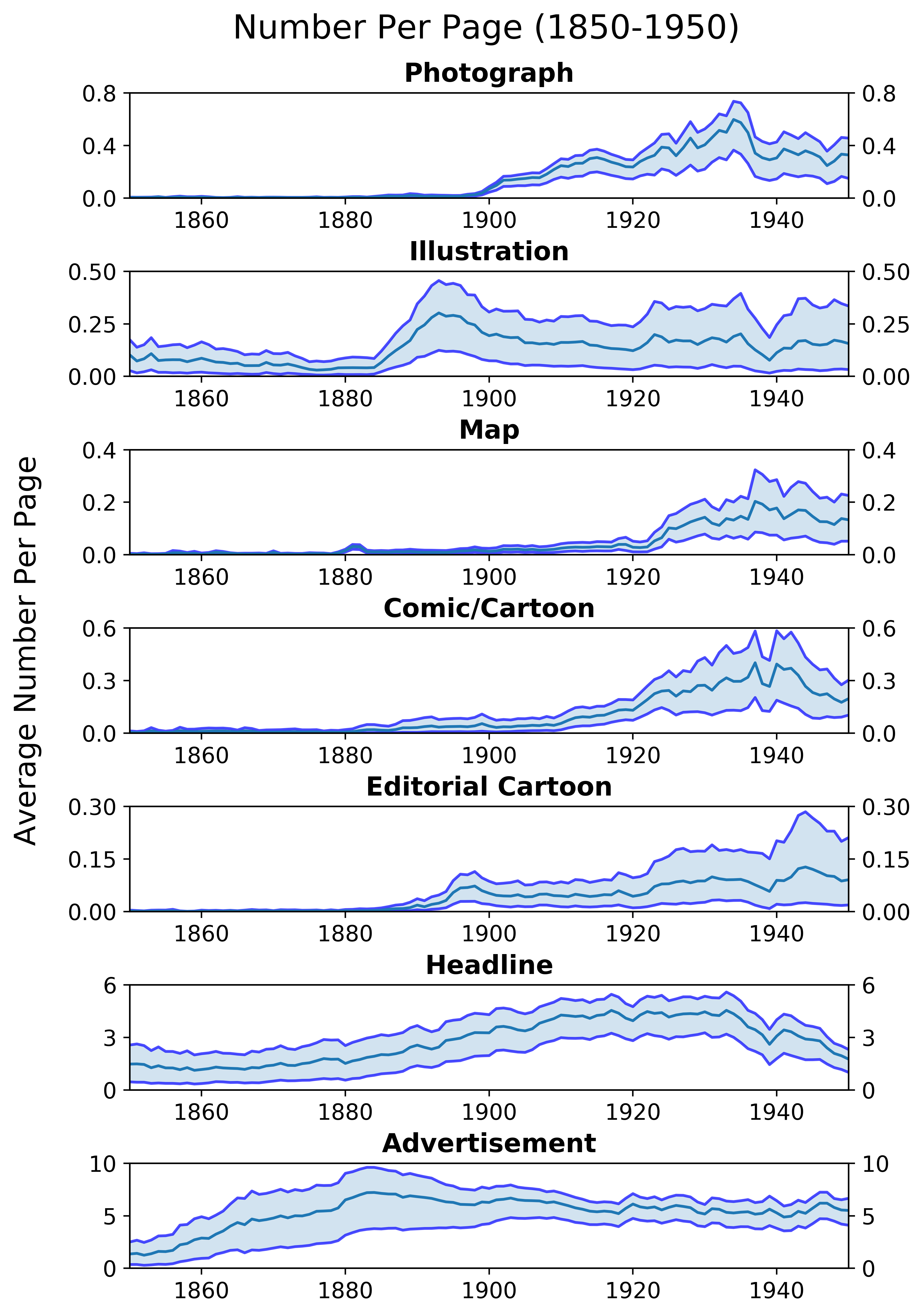}
        \caption{A plot showing the average number of appearances per page of each of the seven classes of visual content, from 1850 to 1950.}
    \end{subfigure}%
    ~ 
    \begin{subfigure}[t]{0.5\textwidth}
        \centering
        \includegraphics[height=5in]{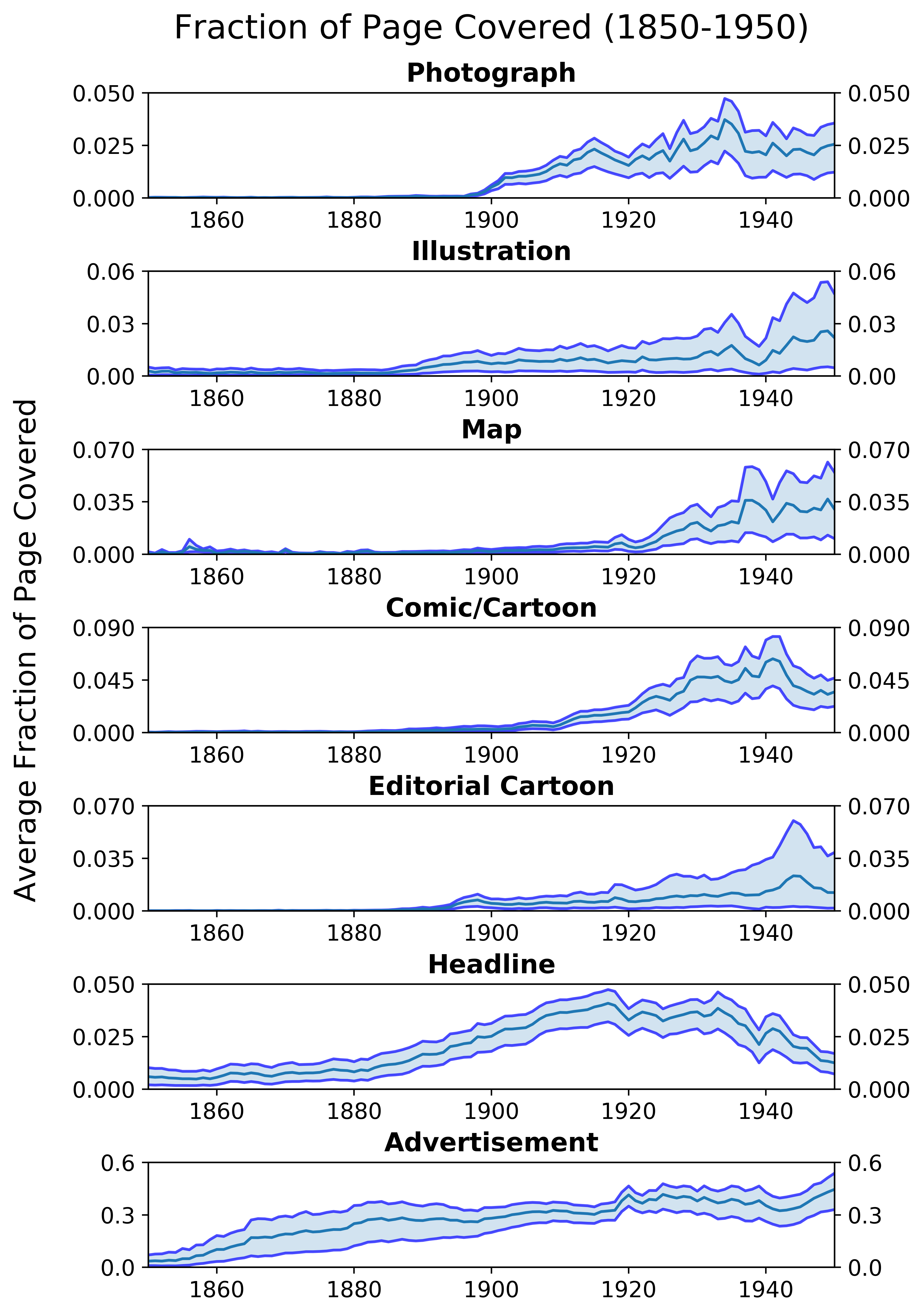}
        \caption{A plot showing the average fraction of each page covered by each of the seven classes of visual content, from 1850 to 1950.}
    \end{subfigure}
    \caption{Multipanel plots visualizing the number of photographs, illustrations, maps, comics, editorial cartoons, headlines, and advertisements in the Newspaper Navigator dataset, derived from 16.3 million historic newspaper pages in Chronicling America.  \textbf{In each plot, the middle line corresponds to a cut of 0.7 on confidence score, and the upper and lower bounds of the confidence interval in light blue correspond to cuts of 0.5 and 0.9, respectively.} Note that the y-axis scales vary per category in both plots. The fraction of each page covered is included because it is a more consistent metric for complicated visual content layouts (such as photo montages and classified ads): predicted bounding boxes can vary greatly in number while still remaining correct and covering the same regions in aggregate.}~\label{fig:datasetviz}
\end{figure*}

In Figure \ref{fig:datasetviz}, we show visualizations of the number of photographs, illustrations, maps, comics, editorial cartoons, headlines, and advertisements in the Newspaper Navigator dataset according to year of publication. These visualizations show the average number of appearances per page of each of the seven classes over time, as well as the average fraction of the page covered by each of the seven classes over time.  As in Table \ref{tab:breakdownextracted}, we show three different cuts. With these visualizations, we can observe trends such as the rise of photographs at the turn of the 20th century and the gradual increase in the amount of page space covered by headlines from 1880 to 1920.

To demonstrate questions that we can begin to answer with the Newspaper Navigator dataset, we have included Figure \ref{fig:maps}, a visualization showing maps of the Civil War identified by searching the visual content for all 278,094 pages published between 1861 and 1865 in the dataset.\footnote{The visualization was created using \cite{collagemaker}.} From this collection alone, researchers can study Civil War battles, the history of cartography, differences in print trends for northern and southern newspapers, and map reproduction patterns (``virality'').

\subsection{Dataset Access and Format}

The Newspaper Navigator dataset can be accessed via the Newspaper Navigator GitHub repository.
We introduce the data format below, but more detailed instructions for use can be found at the webpage. For each processed page, an associated JSON file contains the following metadata:

\begin{itemize}
\itemsep0em
    \item \texttt{filepath [str]}: the path to the image, relative to the Chronicling America file structure.\footnote{For example, see
    \url{https://chroniclingamerica.loc.gov/data/batches/}.}
    \item \texttt{pub\_date [str]}: the publication date of the page, in the format \texttt{YYYY$-$MM$-$DD}.
    \item \texttt{boxes [list:list]}: a list containing the normalized coordinates of predicted boxes, indexed according to [$x_1$, $y_1$, $x_2$, $y_2$], where $(x_1, y_1)$ is the top-left corner of the box relative to the standard image origin, and $(x_2, y_2)$ is the bottom-right corner .
    \item \texttt{scores [list:float]}: a list containing the confidence score associated with each box (only predicted boxes with a confidence score $\ge 0.5$ were retained).
    \item \texttt{pred\_classes [list:int]}: a list containing the predicted class for each box using the following mapping of integers to classes:
    \begin{itemize}
    \itemsep0em 
        \item[--] 0 $\to$ Photograph
        \item[--] 1 $\to$ Illustration
        \item[--] 2 $\to$ Map
        \item[--] 3 $\to$ Comics/Cartoon
        \item[--] 4 $\to$ Editorial Cartoon
        \item[--] 5 $\to$ Headline
        \item[--] 6 $\to$ Advertisement
    \end{itemize}
    \item \texttt{ocr [list:str]}: a list containing the OCR of white space-separated strings identified within each box.
    \item \texttt{visual\_content\_filepaths [list:str]}: a list containing the filepath for each cropped image (except headlines, which were not cropped and saved).
\end{itemize}

Another JSON file with the same file name  with the suffix ``\texttt{\_embeddings}'' includes the image embeddings in the following format; any prediction with a confidence score of less than $0.5$ does not have a corresponding embedding:

\begin{itemize}
\itemsep0em
    \item \texttt{filepath [str]}
    \item \texttt{resnet\_50\_embeddings [list:list]}: a list containing the 2,048-dimensional ResNet-50 embedding for each image (except headlines, for which embeddings were not generated).
    \item \texttt{resnet\_18\_embeddings [list:list]}: a list containing the 512-dimensional ResNet-18 embedding for each image (except headlines, for which embeddings were not generated).
    \item \texttt{visual\_content\_filepaths [list:list]}
\end{itemize}

\subsection{Pre-packaged Datasets}

\begin{figure*}
\centering
  \includegraphics[width=2.0\columnwidth]{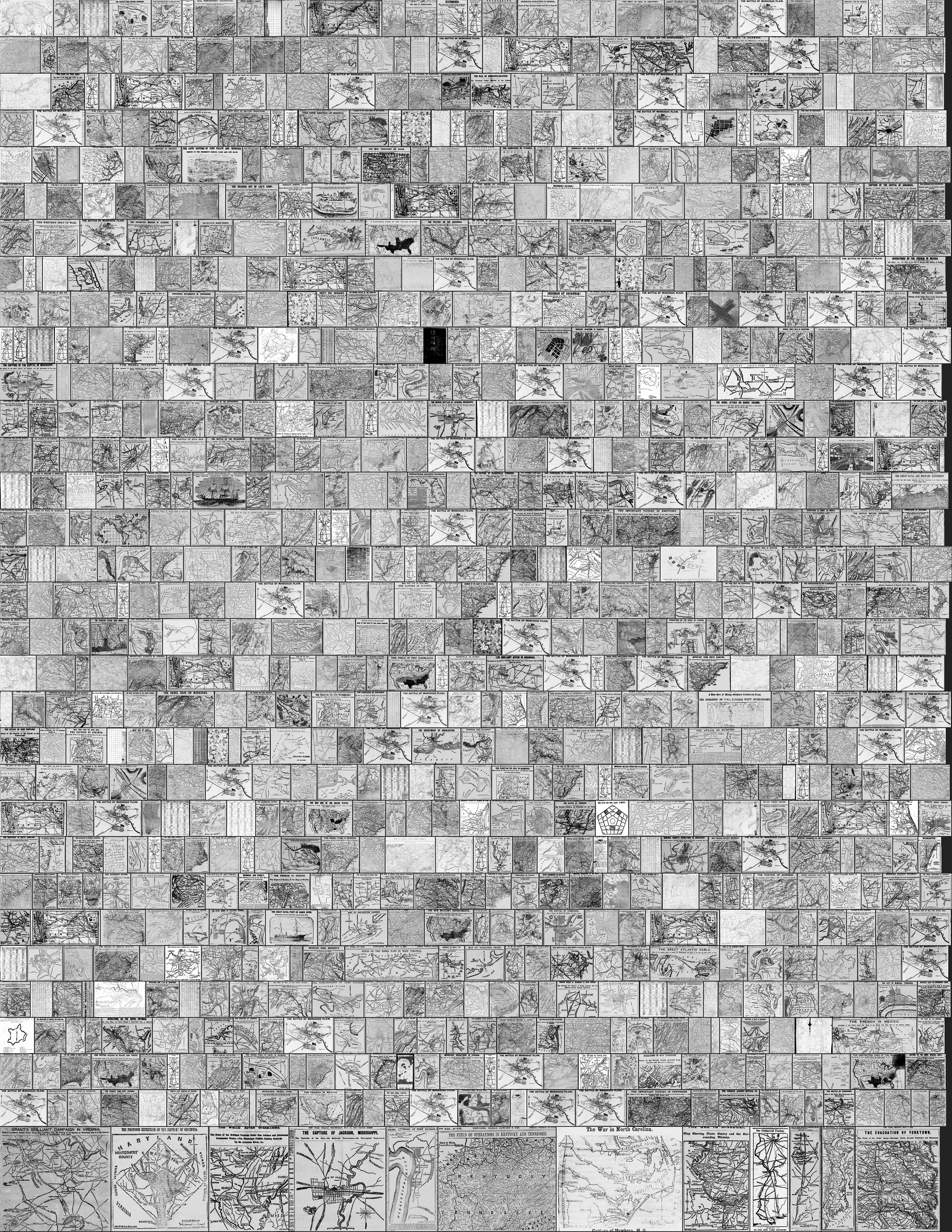}
  \caption{A visualization of the Civil War maps in the Newspaper Navigator dataset, filtered from 278,094 pages published from 1861 through 1865. Note that certain maps appear multiple times, indicating that they were reprinted. Only a small fraction of included images are false positives, suggesting the high performance of the visual content recognition model in the pipeline. This collection of Civil War maps is available as a pre-packaged dataset.}~\label{fig:maps}
\end{figure*}

In order to make the Newspaper Navigator dataset accessible to those without coding experience, we have also packaged smaller datasets derived from the Newspaper Navigator dataset that can be downloaded in bulk from our GitHub repository. These derivative datasets are grouped geographically and temporally and cover both visual content and textual content (machine-readable headlines, captions of visual content, etc.). One such example is the collection of Civil War maps shown in Figure \ref{fig:maps}. We will continue to add derivative datasets as Newspaper Navigator evolves.

\section{Discussion}

\begin{table}
\centering
\begin{tabular}{ccc} 
\multicolumn{3}{c}{\textbf{Performance for 19th Century Newspaper Pages}} \\
\toprule
\centering
\textbf{Category} & \textbf{AP (1850-1875)} & \textbf{AP (1875-1900)} \\
\midrule
Headline & 21.2\% & 51.6\% \\
\hline
Advertisement & 7.3\% & 44.7\% \\
\hline
Illustration & N/A & 36.4\% \\
\hline
One Class & 12.1\% & 48.1\% \\
 \bottomrule
\end{tabular}
\caption{A table showing the average precision (AP) on test sets of 500 annotated pages from 1850 to 1875 and 500 annotated pages from 1875 to 1900. Due to the rarity of the other classes in the labeled data, only headlines, advertisements, and illustrations are included. As in Table \ref{tab:model_performance}, \textit{One Class} refers to the average precision when combining all visual content into one class in order to capture how much error is introduced by the detection of visual content versus the classification.}\label{tab:19th}
\end{table}

\subsection{Generalization to 19th Century Newspaper Pages}\label{sec:generalization}

Given that the visual content recognition model has been trained on World War 1-era newspapers, it is natural to question the generalization ability of the model to 19th century newspapers. Though Figure \ref{fig:datasetviz} reveals trends consistent with intuition, such as the emergence of photographs in historic newspapers at the turn of the 20th century, it is still worthwhile to quantify generalization. To do so, we randomly selected 500 newspaper pages from 1850 to 1875 and 500 newspaper pages from 1875 to 1900 and annotated these pages. In Table \ref{tab:19th}, we present the average precision for headlines, advertisements, and illustrations in the test sets using our annotations as the ground truth (other classes were omitted due to their rarity in the annotated pages). Comparing the results in Table \ref{tab:19th} to the results on the validation data in \ref{tab:model_performance}, we observe a moderate dropoff in performance for pages published between 1875 and 1900, as well as a more major dropoff in performance for pages published between 1850 and 1875. However, the extracted visual content from the pre-1875 pages in the Newspaper Navigator dataset is still of sufficient quality to enable novel analysis, as evidenced by the extracted collection of Civil War maps shown in Figure \ref{fig:maps}.

\subsection{Partnering with Volunteer Crowdsourcing Initiatives}

Our work is also a case study in partnering machine learning projects with volunteer crowdsourcing initiatives, a promising paradigm in which annotators are volunteers who learn about a new topic by participating. With the growing efforts of cultural heritage crowsdsourcing initiatives such as the Library of Congress's By the People \cite{bythepeople}, Smithsonian's Digital Volunteers \cite{smithsonian}, the United States Holocaust Memorial Museum's History Unfolded \cite{historyunfolded}, Zooniverse \cite{zooniverse}, the New York Public Library's Emigrant City \cite{emigrantcity}, The British Library's LibCrowds \cite{libcrowds}, The Living with Machines project \cite{livingwithmachines}, and Trove's newspaper crowdsourcing initiative \cite{trove_crowd}, there are more opportunities than ever to utilize crowdsourced data for machine learning tasks relevant to cultural heritage, from handwriting recognition to botany taxonomic classification \cite{botany}. These partnerships also have the potential to provide insight into project design, decisions, workflows, and the context of the materials for which crowdsourcing contributions are sought. Along with Dielemann \textit{et al}.'s work \cite{galaxyzoo} training a neural network to classify galaxies using crowdsourced data from GalaxyZoo, we hope that our project encourages more machine learning researchers to partner with volunteer crowdsourcing projects, especially to study topics pertinent to cultural heritage.

\section{Conclusion}

In this paper, we have described our pipeline for extracting, categorizing, and captioning visual content, including headlines, photographs, illustrations, maps, comics, editorial cartoons, and advertisements in historic newspapers. We present the Newspaper Navigator dataset, a dataset of these 7 types of extracted visual content from 16.3 million pages from Chronicling America. This is the largest dataset of its kind ever produced. In addition to releasing the Newspaper Navigator dataset, we have released our visual content recognition model for historic newspapers, as well as a new training dataset for this task based on annotations from Beyond Words, the Library of Congress Labs's crowdsourcing initiative for annotating and captioning visual content in World War 1-era newspapers in Chronicling America.  All code has been placed in the public domain for unrestricted re-use.

\section{Future Work}

Future work on the pipeline itself includes improving the visual content recognition model's generalization ability for pre-20th century newspaper pages, especially for the 10.4\% of the pages in the Newspaper Navigator dataset published before 1875. This could be accomplished by finetuning on a more diverse training set, which could be constructed by partnering with another volunteer crowdsourcing initiative such as the Living with Machines project \cite{livingwithmachines}. One could also imagine training an ensemble of visual content recognition models on different date ranges. Given that only 10.4\% of pages in the Newspaper Navigator dataset were published before 1875, it is straightforward to re-run the pipeline with an improved visual content recognition model on these pages.\footnote{This simply requires replacing the model weights file and filtering the pages for processing by date range.} 

To improve the textual content extracted from the OCR, future work includes training an NLP pipeline to correct systematic OCR errors. In the second step of the Beyond Words pipeline, volunteers were asked to correct or enter the OCR that appears over each marked bounding box, resulting in approximately 10,000 corrected textual annotations to date. It is straightforward to construct training pairs of input and output in order to train a supervised model to correct OCR. Other approaches to OCR postprocessing include utilizing existing post-hoc OCR correction pipelines \cite{impresso_ocr, nguyen_2019_b, nguyen_2019_a, datamunging}, all of which could be benchmarked on the aforementioned Beyond Words training pairs.

The future work that excites us most, however, consists of the many ways that the Newspaper Navigator dataset can be used. Our immediate future work consists of building a new search user interface called Newspaper Navigator that will be user tested in order to evaluate new methods of exploratory search. However, future work also includes investigating a range of digital humanities questions. For example, the Viral Texts \cite{ vt5,vt6,vt3,vt1,vt4,vt2} and Oceanic Exchanges \cite{oceanicexchanges} projects have studied text reproduction patterns in 19th century newspapers, including newspapers in Chronicling America; the Newspaper Navigator dataset allows us to study photograph reproduction in 20th century newspapers. In addition, using the headlines in Newspaper Navigator, we can study which news cycles appeared in different regions of the United States at different times. These examples are just a few of many that we hope will be examined with the Newspaper Navigator dataset. We hope to inspire a wide range of digital humanities, public humanities, and creative computing projects.

\section{Acknowledgments}

The authors would like to thank Laurie Allen, Leah Weinryb Grohsgal, Abbey Potter, Robin Butterhof, Tong Wang, Mark Sweeney, and the entire National Digital Newspaper Program staff at the Library of Congress; Molly Hardy at the National Endowment for the Humanities; Stephen Portillo, Daniel Gordon, and Tim Dettmers at the University of Washington; Michael Haley Goldman, Eric Schmalz, and Elliott Wrenn at the United States Holocaust Memorial Museum; and Gabriel Pizzorno at Harvard University for their invaluable advice with this project. Lastly, the authors would like to thank everyone who has contributed to Chronicling America and Beyond Words, without whom none of this work would be possible. 

This material is based upon work supported by the National Science Foundation Graduate Research Fellowship under Grant DGE-1762114, the Library of Congress Innovator-in-Residence Position, and the WRF/Cable Professorship.

%
%
%
%
%
\balance{}

\balance{}

\bibliographystyle{SIGCHI-Reference-Format}
\bibliography{sample}

\end{document}